%% file: main.tex
\documentclass[10pt,journal,compsoc]{IEEEtran}
\usepackage{amsmath,amsfonts}
\usepackage{algorithmic}
\usepackage{algorithm}
\usepackage{array}
\usepackage[caption=false,font=normalsize,labelfont=sf,textfont=sf]{subfig}
\usepackage{textcomp}
\usepackage{stfloats}
\usepackage{url}
\usepackage{verbatim}
\usepackage{graphicx}
\usepackage{cite}
\usepackage{hyperref}

\usepackage{multirow}
\usepackage[table,xcdraw]{xcolor}
\usepackage{tabularx}
\usepackage{booktabs}

\usepackage{tikz}

\newcommand\submittedtext{%
  \footnotesize This work has been submitted to the IEEE for possible publication. Copyright may be transferred without notice, after which this version may no longer be accessible.}

\newcommand\submittednotice{%
\begin{tikzpicture}[remember picture,overlay]
\node[anchor=south,yshift=10pt] at (current page.south) {\fbox{\parbox{\dimexpr0.65\textwidth-\fboxsep-\fboxrule\relax}{\submittedtext}}};
\end{tikzpicture}%
}

\begin{document}

\errorcontextlines=200

\title{The Relationship Between Time and Distance Perception in Egocentric and Discrete Virtual Locomotion (Teleportation)}

\author{Matthias Wölwer, Daniel Zielasko

 \IEEEcompsocitemizethanks{\IEEEcompsocthanksitem M. Wölwer and D. Zielasko are with the University of Trier, Germany. \protect\\
    E-mail: woelwerm@uni-trier.de, zielasko@uni-trier.de}
%\thanks{This paper was produced by the IEEE Publication Technology Group. They are in Piscataway, NJ.}
%\thanks{Manuscript received April 19, 2021; revised August 16, 2021.}
}

%\markboth{IEEE Transactions on Visualization and Computer Graphics}%
%{Wölwer and Zielasko: The Relationship Between Time and Distance
%Perception in Egocentric and Discrete Virtual Locomotion (Teleportation)}


\IEEEtitleabstractindextext{
\begin{abstract}
Traveling distances in the real world inherently involves time, as moving to a desired location is a continuous process.
This temporal component plays a role when estimating the distance covered. 
However, in virtual environments, this relationship is often changed or absent.
Common teleportation techniques enable instantaneous transitions, lacking any temporal element that might aid in distance perception.
Since distances are found to be commonly underestimated in virtual environments, we investigate the influence of time on this misperception, specifically in target-selection-based teleportation interfaces.
Our first experiment explores how introducing a delay proportional to the distance covered by teleportation affects participants' perception of distances, focusing on underestimation, accuracy, and precision.
Participants are required to teleport along a predefined path with varying delays.
A second experiment is designed to determine whether this effect manifests in a more application-specific scenario.
The results indicate a significant reduction in distance underestimation, improving from 27\% to 16.8\% with a delayed teleportation method.
Other sub-scales of distance estimation hardly differ.
Despite targeted adaptations of previous study designs, participants have again found strategies supporting them in estimating distances.
We conclude that time is a factor affecting distance perception and should be considered alongside other factors identified in the literature.
\end{abstract}

}

\maketitle
\IEEEpeerreviewmaketitle

\submittednotice

\begin{comment}
Intro:
motivate:
    - ++context and purpose of the study
    - ++~= why distance perception is important
    - ++the design choichs of the technique
    - ++why continuouse steering was ommited ("only for cybersickness?")
\end{comment}

\IEEEraisesectionheading{\section{Introduction}\label{sec:introduction}}
Target-selection-based teleportation (Teleportation) is one of the most common methods of locomotion in VR applications.
Since it is easy to use and not being cybersickness-inducing it is a solid choice for numerous applications.
VR applications are becoming more relevant in areas of research and training, as they allow the user's body and environment to be manipulated in ways that are difficult to reproduce in the real world \cite{CreemRegehr2022perceivingDist}.
Especially for applications, studying human perception but also applications like viewing virtual architecture \cite{Frost} or museums \cite{Zielasko2020virtMuseum} that convey the size of historical places, one important cornerstone when trying to transfer results and impressions gained in the virtual world to the real world, is correctly assessing distances.
Research shows that distances in VR are commonly underestimated, only averaging around 71~\% to 73~\% of the actual distances \cite{Waller2008CorrectingDistances, Renner2013}.
Similar underestimations are observed when estimating distances using teleportation \cite{Keil}.
Factors contributing to this underestimation are numerous \cite{CreemRegehr2022perceivingDist} but so far no conclusive reason can be named.
Looking at locomotion, one potential reason for the underestimation of distances that has been neglected in research so far is the time that passes when moving the user from their initial to their desired point in the virtual world.
To close this gap, we aim to shed light on distance perception under the influence of time during locomotion.
While an analysis of a temporal component seems intuitive for locomotion methods with a continuous transition like steering we investigate methods with discrete transitions, as teleportation is the de facto standard locomotion method in many applications due to its reduced influence on cybersickness.
Specifically, we investigate how distance perception changes when using teleportation with different temporal aspects.

Teleportation consists of two main phases, which are a target specification (selection) and transition \cite{Weissker2018Jumping}. 
While the change of the user's position happens instantaneously and is the namesake for the methods, i.e., usually without any continuous movement, the target specification process itself does not necessarily need to be discrete, nor does the transition itself need to happen instantaneously after the target is specified. 
A simple fade in and out to black, between the change of position, for example, is very common. 
Thus, it is evident that time is an (often unnoticed) parameter in teleport interfaces. 
Therefore, we want to tackle the following research questions with this work:

\textbf{RQ1 Is there an impact of delayed teleportation on the perception of virtually traveled distance?}
In the following, we will attempt to answer this question by breaking it down further:
\textbf{RQ1.1} Is there a difference between a distance proportional delay that matches human walking speed and a delay that is proportional but faster than this? 
\textbf{RQ1.2} Is there a difference when the target specification process is discrete vs continuous?

In order to answer the research questions we evaluate different delays of the teleportation process.
As a baseline for RQ1, we use a standard target-selection-based teleport interface, also referred to as Point \& Teleport, where the user discretely selects a target point and is transitioned instantaneously after a quick fade-to-black animation.
To evaluate RQ1.1 we delay the teleportation process by the speed of human walking through a simplified avatar, that moves from the initial position to the target position and triggers an instantaneous translation of the user after reaching the target.
The use of walking speed is motivated by aiming to make results gathered in the virtual world transferable into the real world where walking speed is the most familiar speed of travel for humans.
The same method but with increased walking speed is used for comparison.
To answer RQ1.2 we move the delay caused by the avatar movement from the pre-travel information phase to the target-specification phase \cite{Weissker2018Jumping}.

Finally, applications that allow the viewing of virtual architecture (c.f. \cite{Frost, Zielasko2020virtMuseum}) motivate us to examine if possible effects from RQ1 have an impact on the perception of room sizes that users can walk through.
Thus we further ask: 
\textbf{RQ2 Is a potential effect also visible in applications, such as room size perception?}

\section{Related Work}\label{sec_rw}
To the best of our knowledge, there is no prior work explicitly investigating the effects of time on distance estimation in discrete virtual locomotion techniques.
However, time is often implicitly and in various manifestations a factor in different stages of the teleportation process, which we want to discuss in Section~\ref{sec_rw_time-in-locomotion}.
Furthermore, distance estimation in the context of VR and locomotion is an established field of research, which we want to touch on in Section~\ref{sec_rw_distance-estimation}.

\subsection{Temporal Aspects in Teleportation\label{sec_rw_time-in-locomotion}}
Weissker et al. propose dividing the teleportation process into $4$ stages \cite{Weissker2018Jumping}: target specification, pre-travel information, transition, and post-travel feedback. 
We would like to follow this model and, in the following, work out how in prior research the different stages consider or intentionally implement the factor of time.

\textbf{Target Specification:}
The selection of the teleportation target is primarily implemented through pointing in most cases, hence the commonly used term 'Point \& Teleport. 
Although there are various implementations (e.g., ray vs. parabola), selecting a target point itself is always a temporal factor as it requires time to point the controller (or differing methods of input) toward the desired target point.

Besides this unavoidable temporal factor, implementations are adding a constant dwell time as part of the selection: 
We find Bolte et al. adding $0.5~s$ \cite{Bolte2011Jumper} and Bozgeyikli et al. with $2~s$ \cite{Bozgeyikli2016PointAndTP}.
Methods that employ some kind of cursor steered toward the target location (iterative process) introduce a delay proportional to the traveled distance \cite{Griffin2019OutofBody, Drogemuller2020NavTech}.
Griffin et al. utilize a speed of $6~m/s$ \cite{Griffin2018Handybusy} and Weissker et al. allow for a maximum speed of $3~m/s$ by explicit selection from a continuous range \cite{Weissker2023HighGround}.

\textbf{Pre-Travel Information:}
Feedback extending beyond the selection process is rarely encountered in interfaces.
Bolte et al. visualize a target indicator that grows over $2~s$ before the transition \cite{Bolte2011Jumper} and Rupp et. al
animate changes between the selection and the current frame \cite{Rupp2023Tenet}.
Cmentowski et al. use an avatar that moves to the target position for a delay that is proportional to the covered distance 
\cite{Cmentowski2019Outstanding}.

\textbf{Transition:}
During the transition phase the user effectively ``moves'' from the starting point to the destination point.
Unfortunately, in about $30\%$ of the papers we investigate the exact type of a transition is not specified, which matches observations of other research in this field \cite{zielasko2023crises}. 
We assume that one reason for the missing information often is that the transition occurs instantaneously, i.e., ``nothing'' happens.
Even if we assume an instantaneous transition only in cases where it is explicitly specified, the number of such instances remains substantial \cite{Bowman1997TravelTechnique, Christou2017SteeringVSTeleport, Langbehn2018RoomScale, Coomer2018FourLoco, Griffin2018Handybusy, Clifton2019SteeringTeleport, Paris2019VideoGame, Boletsis.2017, Cherep2020CognitiveImplications, Bozgeyikli2016PointAndTP, Griffin2019OutofBody, Weissker2018Jumping, Weissker2019MultiRay, Weissker2020GettingThere, Sayyad2020WideArea, Lai2021CognitiveLoads, Bimberg2021VirtRotations, Kelly2022ReplicatingLabStudy, Hombeck2023TellMeWhereToGo, Liu2018Walking, Sayyad2020WideAreaWalking, Kelly2022CompareLabAndAtHome}.
Another constant but distinct from zero transition, is a fade-to-black effect between the two viewpoints.
Lindal et al. set the total time to $1.6~s$ \cite{Lindal2018Comparison}, Drogemuller et al. to $1.0~s$ \cite{Drogemuller2020NavTech}, Rahimi et al. $1.5~s$ \cite{Rahimi2020SceneTransitions}, and Freitag et al. do not specify the exact time \cite{Freitag2018Exploration}.
We would also consider walking through a portal as a constant time manipulation, as the distance traveled through the portal in no known implementation correlates with the actual distance covered \cite{Liu2018Walking, Freitag.2014, Feld2023transitionLBW}.
Most animated view-point transitions tend to exhibit a duration proportional to the distance covered:
Bolte et al. move the user with $5.56~m/s$ \cite{Bolte2011Jumper}, Bhandari et al. and Adhikari et al. $10~m/s$ \cite{Bhandari2018Dash, Adhikari2022hyperjump}, Habgood  $5~m/s$ \cite{Habgood.2017}, Lai et al. $1.44~m/s$ \cite{Lai2021CognitiveLoads}, and Rahimi et al. use $10$, $25$ and $50~m/s$ \cite{Rahimi2020SceneTransitions}.
The latter work, furthermore, investigates the impact of $3$-$5$ intermediate jumps, however, this is not related to the distance covered.
Some works do not provide precise details on movement speeds \cite{Chen2022UrbanRama, Cmentowski2019Outstanding}.
Finally, Lee et al. employ a variable speed that depends on the immediate environment \cite{Lee2023ViewportTransitions}, therefore, it does not behave proportionally to the distance.

\textbf{Post Travel Feedback:}
For the post-travel feedback, we only find Bimberg et al. \cite{Bimberg2021VirtRotations} visualizing an out-fading rotation axis over $0.5~s$.

The works of Rahimi et al. \cite{Rahimi2020SceneTransitions} and Lai et al. \cite{Lai2021CognitiveLoads} implicitly deal with different implementations of time in teleportation (see above). 
However, their focus is limited to aspects of spatial awareness, cybersickness, cognitive loads, and usability rather than distance perception.

\subsection{Distance Perception}\label{sec_rw_distance-estimation}
Existing studies on distance perception in VR reveal that when using Head-Mounted Displays (HMDs), distances in virtual environments are typically underestimated.
A meta-study by Waller and Richardson \cite{Waller2008CorrectingDistances} reveals that estimated distances in virtual environments average only $71\%$ of the actual distance.
Similarly, Renner et al. \cite{Renner2013} find that distance estimation averages only about $73\%$ of the actual distance. 
Messing and Durgin \cite{Messing2005distPerc} find distances to average $77\%$.
Estimations in a non-virtual environment result in an accuracy of $96\%$ \cite{Messing2005distPerc} and go up to $99.9\%$ \cite{Waller2008CorrectingDistances}.

Factors affecting distance perception include technical (hardware and parameters), compositional (environmental features), human (psychological traits), and measurement-related factors \cite{Renner2013}.
These factors collectively impact the accuracy of distance perception and estimation.
Moreover, research by Creem-Regehr et al. \cite{CreemRegehr2022perceivingDist} identifies cues for adjusting size perception in virtual environments.
Cues to calibrate the perception of scale in virtual environments are influenced by user-related experiences like familiarity with the virtual environment, virtual avatars, maintaining correct eye level, and experience of movement.
On a technological side the field of view and HMDs weight influence perception \cite{CreemRegehr2022perceivingDist, Messing2005distPerc, Kelly2023DistPercMetaAnaysis, Masnadi2022EffectsOfFOV}.
Furthermore distances are perceived differently when viewing objects right in front of the user versus on the side in the virtual environment  \cite{Peillard2019ObjectsLookFarther}.

Renner et al. \cite{Renner2013} also provide an overview of common measurement techniques used to quantify distance perception. 
Verbal estimation, adjustment tasks, and blind walking are enumerated.
Jamiy and Marsh \cite{jamiy-de-survey} present a similar categorization of methods including matching tasks, verbal reports, bisection tasks, and blind walking.
In addition, they present experiments highlighting the limitations of these methods, with blind walking providing imprecise estimations beyond $20$ meters, verbal reports tending to underestimate, and the accuracy of estimations using the bisection task varying based on the virtual environment.
Furthermore, Bruder et al. \cite{Bruder2015DistEstim} study distance perception in immersive projection environments, revealing that screen distance and parallax significantly impact distance estimation, emphasizing the importance of relative positions of users and virtual objects.

A study by Keil et al. \cite{Keil} explores how different methods of locomotion in VR affect distance estimation.
Two locomotion methods are compared regarding distance estimation: instantaneous teleportation with target specification via pointing a parabolic ray and a hand-directed steering method allowing forward and backward movement with a speed of $1.39~m/s$.
Findings show that distance estimates were more accurate when using the teleportation method. 
Keil et al \cite{Keil}, however, note that counting strategies for estimating distance, either by counting seconds of movement during steering or counting equal-sized teleportation jumps, can influence estimations, especially after participants were trained with a reference size.
This experiment by Keil et al. \cite{Keil} is particularly relevant to the present work as it represents the only known investigation of the effects of teleportation on distance estimation.

\section{Method\label{sec_method}}

\begin{figure}
    \centering
    \includegraphics[width=1.0\columnwidth]{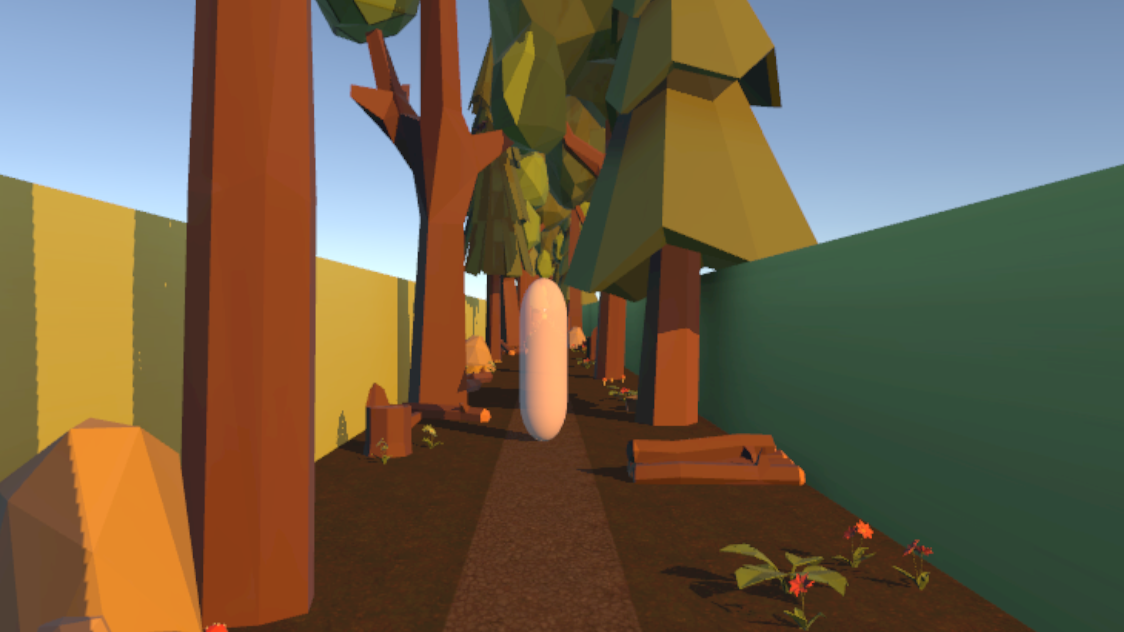}
    \vspace{-1.5em}
    \caption{Virtual environment of the distance estimation and distance repetition task with a capsule representing the users' movement on the path when using the delayed teleportation methods.}
    \label{img_distPerc_virtEnv_self}
\end{figure}

To address the research questions we manipulate the temporal behavior of different stages in a Point \& Teleport interface (see Section~\ref{sec_discPerc_tpm}).
To clarify the central parameters of the implementation regarding its temporal behavior, we conduct a preliminary study as a first step and describe this in Section~\ref{sec_prestudy}.
Subsequently, we conduct two empirical studies.
The study design and our hypotheses before the studies were preregistered online \footnote{\url{https://doi.org/10.17605/OSF.IO/VEK3W}}.
The study received approval from the local ethics board.
The objective of the first study is to assess the impact of temporally delayed teleportation on distance perception.
To isolate potential effects we extend an existing task design from Keil et al. \cite{Keil} in an artificial but very controlled distance estimation task (see Section~\ref{sec_study1}).
In a second study, we try to reveal potential effects in a more natural application of distance estimation, i.e., the size perception of rooms much like it is of interest during an architectural walkthrough, apartment viewing, or furniture arrangement (see Section~\ref{sec_study2}).

\subsection{Hypotheses}\label{sec_discPerc_hypo}
While the realm of temporal influence on distance perception during teleportation remains unexplored, the findings of related studies allow for the formulation of hypotheses:
\begin{itemize}
    \setlength\itemsep{-.2em}
    \item[\textbf{H1}] A proportional delay of the teleportation (matching the time it would take a human with average walking speed to cover the distance of teleportation) will increase the performance in distance estimation when compared to classic teleportation without a delay.
    \item[\textbf{H2}] A delay of the teleportation oriented on human walking speed will lead to a higher estimation of distance when compared to a still proportional but faster delay.
    \item[\textbf{H3}] A continuous target specification process (oriented on walking speed) will differ in performance in distance estimation when compared to a discrete and delayed (oriented on walking speed) specification.
    \item[\textbf{H4}] A proportional delay of the teleportation will increase the performance in the estimation of room sizes compared to classic teleportation without a delay.
\end{itemize} 

A rationale for the hypotheses primarily arises from psychological insights.
The hypothesis that a proportionally delayed teleportation improves distance estimation compared to instantaneous teleportation (\textbf{H1}) is grounded in various research findings.
Various studies indicate, distances in virtual space are underestimated by approximately $27\%$ \cite{Waller2008CorrectingDistances, Renner2013}. 
The rationale for \textbf{H1} follows from this knowledge, considering the \textit{Tau Effect}.
The Tau Effect can be described as the impact of time on distance perception, wherein differences in the duration of a time unit under the same conditions directly affect spatial estimation \cite[p.14]{Fraisse1984}.
Helson and King \cite{tauEffect_Helson1931} observe in an estimation experiment that the stimulation of three equidistant points $p1$, $p2$, and $p3$ at time points $t1$, $t2$, and $t3$, respectively, with different durations $||t2 - t1|| > ||t3 - t2||$ leads participants to perceive the distance between $p1$ and $p2$ as longer than between $p2$ and $p3$ \cite{BruderTime}.
Thus, the perception of a longer duration results in a longer perception of distance.
Based on the general underestimation in virtual space, this suggests that a delayed time to reach the destination point leads to a longer perception of the distance traveled. 
We hypothesize, that it thereby improves distance estimation performance by reducing underestimation.
Hence, we expect the underestimation observed in virtual environments of around $27 \%$ to come closer to the underestimation of $4 \%$ \cite{Messing2005distPerc} observed in real-world environments.

The rationale for \textbf{H2} is not only based on an analogous application of the Tau Effect, which suggests that a shorter delay causes a correspondingly shorter distance perception compared to a longer delay but is also supported by insights from the work of Frenz and Lappe \cite{Frenz2005}.
The authors investigate egocentric simulated movement and find, in the context of two experiments, that the duration of movement influences distance perception.
According to the work, distances that are reached in less than $3$ seconds are more likely to be underestimated, and distances that require a movement longer than this are more likely to be overestimated.

\textbf{H3} is based on the assumption that continuous target selection differs in its application in various factors from discontinuous target selection.
Continuous target selection requires a more precise focus on the distance to be covered than a simple selection of targets at a certain distance and would therefore justify an improvement in distance perception.
At the same time, it can be assumed that continuous target selection leads to an increased focus on the use of the teleportation method itself, thereby reducing attention to distance perception.
In this case, a poorer estimation is to be expected.
It is for these reasons that we do not expect a particular direction for \textbf{H3}, as we do not know how both considerations will influence the performance in distance perception.

\textbf{H4} follows from the argumentation of H1. 
If distance perception is improved, it can be assumed that this also has positive effects on the perception of space that a user traverses. Thus, we expect a better performance in the perception of room sizes.

\subsection{Interfaces}\label{sec_discPerc_tpm}

\begin{figure}
    \centering
    \includegraphics[width=1.0\columnwidth]{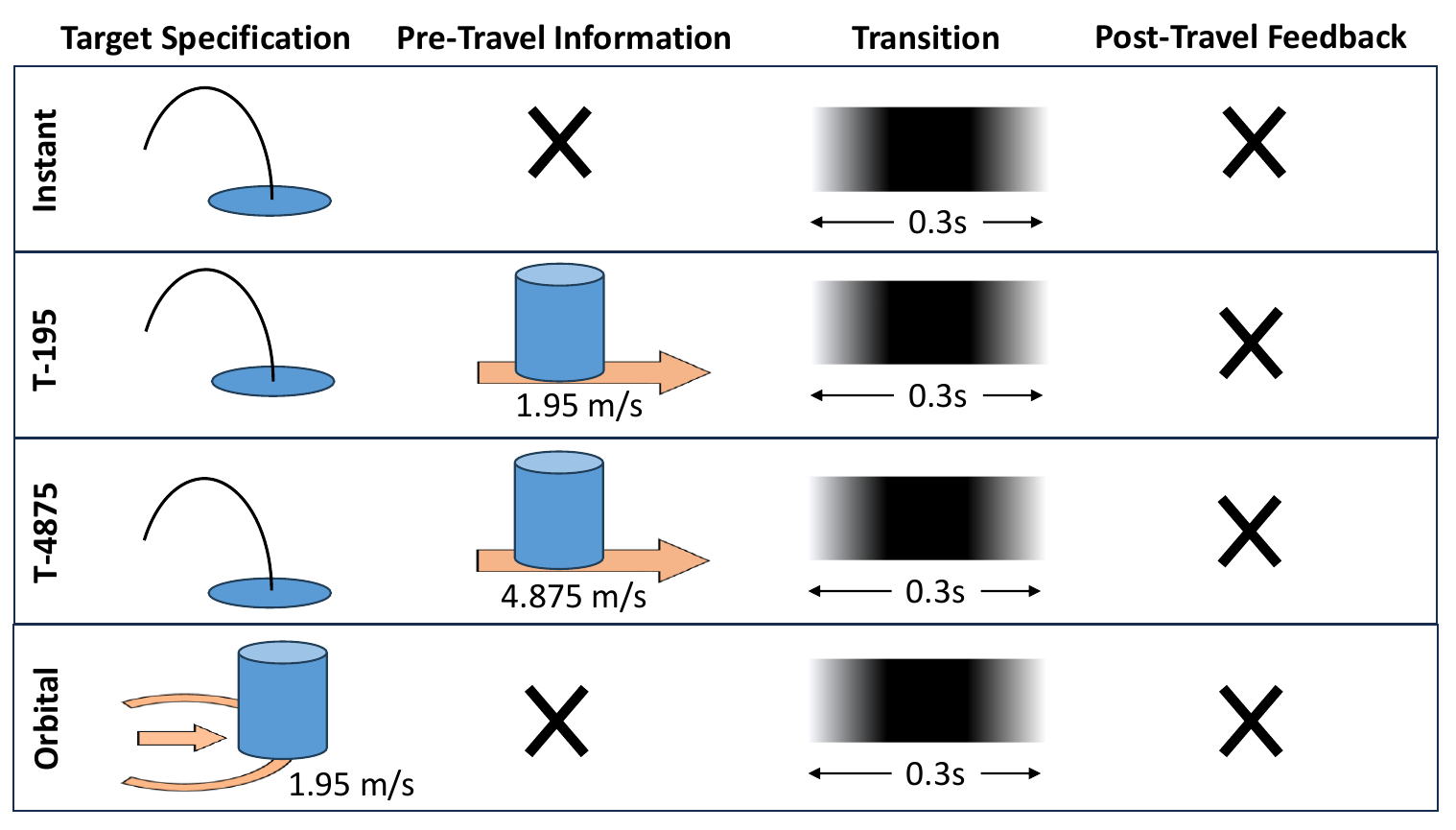}
    \vspace{-1.5em}
    \caption{Individual components of the teleportation process for all interfaces used in our experiments.}
    \label{img_interfaces}
\end{figure}
To evaluate the formulated hypotheses, a total of four teleportation methods are developed, modulating time differently.
As a baseline for hypothesis H1, we use a common Point \& Teleport interface that allows users to discretely select a target point and transition instantaneously.
A second method is developed to compare the instant teleportation with one that works similarly in terms of discrete target selection but introduces a delay that is proportional to the distance covered by the teleport.
To evaluate hypothesis H2, we use this proportionally delayed method and compare it to a third method that works identically but with a delay that is still proportional but faster.
Lastly, we develop a method to compare one of the delayed methods (that has a discrete target selection) with a method that is equally delayed but implements this delay via a continuous target selection.

To introduce a delay we decided on two central parameters: first, as a distance-dependent (proportional) delay, we use the speed of human walking, second, we add a fade-to-black transition lasting a constant duration.
The fade-to-black transition is implemented in all teleportation methods and functions as a visualizer for the transition process.
To implement walking speed as a proportional delay, different phases of the teleportation process can be used.
As our hypothesis H3 requires an explicit comparison of a delay implemented in the target specification phase and a different phase, the choice remains between the pre-travel information and the transition phase.
The first thing that likely comes to mind is manipulating the transition phase. There are two possibilities here: 
1) One can change the instant transition of teleportation into a continuous one. 
However, there is a risk of cybersickness, so we decided against this solution. 
2) One can manipulate the duration of the fade transition. 
We also decided against this solution because we do not want the user to wait for their transition with a black screen, nor do we want a long fading transition that would interfere with subsequent interactions.
Thus, we actively decided against manipulating the transition phase between conditions.
Instead, a virtual representative is chosen to visualize a delay during the target specification or the pre-travel information phase.
This leaves the user stationary while exocentrically watching their representative move toward the target.
The use of a delay oriented on walking speed and a fade-to-black transition is further motivated and suitable parameter choices are determined through a preliminary study described in Section~\ref{sec_prestudy}.
The results show that exocentric walking speed is perceived as $1.95~m/s$ and a duration of $0.3~s$ is preferred for a fade-to-black animation.

Following the schema of Weissker et al. \cite{Weissker2018Jumping}, we describe the teleportation methods along the four phases target specification, pre-travel information, transition, and post-travel feedback.
Before elaborating on the individual teleportation methods, we give a brief overview of the possible implementations of these phases.
Target specification is implemented either \textit{discretely} (no delay) or \textit{continuously} (delayed proportionally to the distance covered).
The pre-travel information phase is solved in one of three ways. 
Either with \textit{no delay}, with a \textit{delay proportional to the distance oriented on the speed of human walking}, and thirdly again a \textit{delay proportional to the distance but oriented on an increased walking speed}.
The transition phase is always the same fade-to-black animation for all methods.
Lastly, the post-travel feedback phase is always omitted.
Fig.~\ref{img_interfaces} visualizes the individual components for our final interfaces.

\subsubsection{Instant Teleportation}
The first and fundamental teleportation method moves the user instantaneously (except for the duration of a fade-to-black animation).
\textbf{\textit{Target specification}}: Users hold a controller from which a parabolic ray emerges when a button is pressed.
Fink et al. \cite{Funk2019Accuracy} suggest enhancing the parabola with a component enabling orientation specification.
However, for our purposes, a standard parabola is sufficient and avoids the small delay introduced by this additional specification.
With this, a target point at a maximum distance of $11~m$ from the current location can be selected.
This distance corresponds to the maximum teleportation range chosen by Bhandari et al. \cite{Bhandari2018Dash} and is used similarly in other implementations. 
Unfortunately, research papers rarely specify the maximum distance.
Further exceptions are Freitag et al. using $7~m$ \cite{Freitag2018Exploration}, Simeone et al. $10~m$ \cite{Simeone2020SpaceBender}, and Weissker et al. $180~m$ \cite{Weissker2018Jumping}.
The target point is indicated by a flat circle on the ground. 
Releasing the button selects the currently marked point as the target.
The \textbf{\textit{pre-travel information}} phase is omitted.
\textbf{\textit{Transition}}: The screen fades to black and the user is set to the target position.
The screen then fades back to clear vision.
The total duration of this animation is $0.3~s$.
The \textbf{\textit{post-travel feedback}} phase is omitted.
Thus, the teleportation method includes a constant delay of $\Delta t = 0.3~s$.
In the following text, we refer to this teleportation method as \textit{instant teleportation}, \textit{Instant}, or \textit{IN} for brevity.

\subsubsection{Teleportation at Walking Speed}
The second method integrates a distance-proportional delay, with its speed aligned with the perceived walking speed of an average physically unimpaired person.
\textbf{\textit{Target specification}}: Users discretely select a target point as described for the instant teleportation method.
\textbf{\textit{Pre-travel information}}: After selecting a target point, a body-sized capsule (see Fig.~\ref{img_distPerc_virtEnv_self}), intended to represent an avatar of the user, moves from the current location to the selected target point in a straight line with a constant speed of $1.95~m/s$.
Once the capsule reaches the target point, the pre-travel information phase ends.
\textbf{\textit{Transition}}:
A fade-to-black animation is identical to the instant method.
The \textbf{\textit{post-travel feedback}} phase is omitted.
For a distance $d$ (in meters) from the starting point to the target point, the delay of this method is given by $\Delta t = \frac{d}{1,95}~s + 0.3~s$.
In the following, this teleportation method is referred to as \textit{teleportation at walking speed} or briefly in reference to its speed as \textit{T-195} or \textit{T1}.

\subsubsection{Teleportation at Increased Walking Speed}
The functionality of the third teleportation method is identical to teleportation at walking speed.
The only difference is that the speed of the capsule during the \textbf{pre-travel information} phase has been increased by a factor of $2.5$.
We chose the factor of $2.5$ as we found it suitable in prior internal testing having two main advantages: 1) it is noticeably faster than the original delay while 2) not being too fast and therefore too similar to the instant transition on short distances.
For a distance $d$ (in meters) from the starting point to the target point, the delay is given by $\Delta t = \frac{d}{2.5 \cdot 1.95}~s + 0.3~s = \frac{d}{4.875}~s + 0.3~s$.
In the following, this method is referred to as \textit{teleportation at increased walking speed}, \textit{T-4875} or \textit{T4}.

\subsubsection{Teleportation with Continuous Target Selection}
\textbf{\textit{Target specification}}: Unlike the target specification phase of the previous methods, here the target point is determined through a continuous process.
In the previous methods, the user discretely selects a target point and then the delay is introduced by our virtual representative which moves in a straight line without stopping or taking any detours to the target point and triggers the transition upon arrival.
To replicate this temporal component as closely as possible, we decided to not allow any kind of stopping or taking a detour during the continuous selection as well.
We therefore do not implement the continuous target selection through steering of the virtual representative as one might expect but through a more limited approach.
When a button is pressed, a capsule starts moving forward from the user's position.
To allow moving the capsule to the left and right within the bound of the path (as one could do with the other teleportation methods), we place the capsule on an imaginary circle around the user.
While the button is pressed this circle continuously enlarges without a maximum limit and pointing the controller in a direction places the capsule on the intersection with the circle.
The speed of enlargement (or forward movement of the capsule) was again chosen as $1.95~m/s$.
Releasing the button selects the point where the capsule is located at that time as the target point.
This allows the user to freely select a target point as the discrete methods would allow.
Reducing the radius of the circle or temporarily stopping without selecting a target point is not possible, as it would not be possible for the discrete methods to stop the capsule while moving.
Any kind of confirmation or possibility of restart would prolong the temporal component and therefore make it incomparable to the discrete methods.
The \textbf{\textit{pre-travel information}} phase is omitted.
\textbf{\textit{Transition}}: A fade-to-black animation identical to all other methods.
The \textbf{\textit{post-travel feedback}} phase is omitted.
This results in a distance-proportional delay with continuous target selection.
For a distance $d$ (in meters) from the starting point to the target point, the delay is given by $\Delta t = \frac{d}{1.95} + 0.3$.
In the following, this teleportation method is primarily referred to as the \textit{Orbital method}, or \textit{OR} due to the capsule's orbit around the user.

\subsection{Prestudy: Fade-to-Black \& Walking Speed\label{sec_prestudy}}
To evaluate the influence of temporal modulations on distance perception, different parameters regarding the duration of a teleportation method need to be established.
As Creem-Regher et al. \cite{CreemRegehr2022perceivingDist} describe, one of the main reasons for requiring correct distance perception is the transferability of results, from the virtual to the real world.
It is therefore reasonable to base the choice of a time manipulation on aspects that are familiar to humans from their natural environment.
The duration of a delay is therefore sensibly chosen if it is proportional to the distance covered.
It can be argued that a person's walking speed is a natural choice since it has everyday applications.
Several works investigate whether the average walking speed in the real world is also perceived as this in the virtual world. 
Bohannon \cite{Bohannon1997WalkingSpeed} shows that the mean comfortable walking speed lies in the range of $1.27~m/s$ to  $1.46~m/s$ with a maximum recorded speed of $2.53 ~m/s$,  Janeh et al. \cite{Janeh2017WalkingInVR} and Fink et al. \cite{Fink2007ObstacleAvoidance} find that users physical walking speed generally decreases in virtual environments.
Other than this, one could also consider calibrating walking speed based on participants walking speed in advance of the study.
This idea and further ideas proposed in the literature, however, only consider egocentrically perceived walking speed.
In our teleportation technique, the user stays still and watches a virtual representative move instead of moving themselves.
We, therefore, see the need to assess how people perceive walking speed when observing a virtual representation from an outside perspective (exocentrically), rather than the egocentric perspective used in previous studies.
Another aspect is the visualization of a transition when using teleportation.
Instead of abruptly transferring users to a new, selected target position, a quick fade helps with a smoother transition between positions \cite{Bozgeyikli2016PointAndTP} and at the same time provides the user with feedback that a movement has taken place, which is not given in the case of discrete teleportation \cite{Riecke.2021}.
A fade-to-black animation is often used for this purpose.
The duration of this transition method represents another integration of time into the teleportation process.
However, this duration is not standardized and has been modeled differently within the research.
Examples are Líndal et al. \cite{Lindal2018Comparison}, with a duration of $1.6~s$ or Rahimi et al. \cite{Rahimi2020SceneTransitions} with a duration of $1.5~s$.
This motivated us to run a preliminary study on the evaluation of these parameters, to answer the following research questions:
\textbf{P-RQ1:} When exocentrically observing a virtual representative moving at different speeds, which speed is perceived as equivalent to the average human walking speed?
\textbf{P-RQ2:} What duration is most preferred for a fade-to-black transition when using Point \& Teleport?
The apparatus for this study is the same as the one for the main Study I (Section \ref{sec_distPerc_apparatus}).

\subsubsection{Procedure \& Task}
In the beginning, participants are asked to indicate their gender, age, and experience level with VR. 
Participants are then immersed in a virtual environment similar to the one used in Study I (see Fig.~\ref{img_distPerc_virtEnv_self}) and can freely explore it using a target selection-based teleportation method. 
The teleportation method differs depending on the question to be answered. 
In both cases, the target specification is solved by a parabolic ray with a maximum range of $11~m$.
To evaluate question \textbf{P-RQ1}, the teleportation method is supplemented with a moving capsule that is exactly the height of the participant and has a horizontal diameter of $0.8~m$  (see Fig.~\ref{img_distPerc_virtEnv_self}) during the pre-travel information phase.
The capsule moves from the user's position to the selected target point and triggers an immediate transition upon arrival. 
Participants are asked to manipulate the capsule's movement speed until it corresponded to their perceived walking speed.
The speed of movement can be adjusted with an initial speed of $1.3~m/s$  and increments of $\pm 50\%$. 
The initial speed is chosen according to the average human walking speed.
Once participants confirm the speed they perceive as walking speed, the application is terminated. 
We are aware that users potentially have different understandings of walking speed, e.g., when walking through a park or when trying to catch a bus.
Such interpretation of the meaning of walking speed is left to the individual participant but the forest-themed environment and absence of an actual task other than parameter specification likely have an impact.

Following the completion of this task, the second application for manipulating the duration of the fade-to-black transition is started (\textbf{P-RQ2}). 
For this task, the moving capsule is removed and a fade-to-black animation is added during the transition phase.
Participants are again asked to move freely and engage in adjusting the duration of the animation. 
The duration is initially set to $0.4~s$, as this was perceived best in an internal test phase.
The $0.4s$ consists of a fade-out and a fade-in taking up half of the time each.
The duration can be adjusted in steps of $\pm 0.1~s$ with a minimum duration of $0.1s$. 
Once a preferred duration is set, the application is terminated, and the survey ends.

\subsubsection{Participants}
Within the possibilities of a prestudy, we tried to keep the sample as diverse as possible, taking into account factors such as age, experience with VR, and gender.
This led to a total of 9 participants ($5$ male, $4$ female) between the ages of $25$ and $67$ years ($M=39.5$ years, $SD = 15.9$) being recruited for the study. 
All participants were reached through the personal and professional networks of the experimenter.
A normal or corrected vision was a prerequisite for participation in the study. 
There was no compensation for participation. 
Of the nine participants, six reported having a lot of experience with virtual reality, one person reported occasional use, and two people reported no experience.

\subsubsection{Results}
In response to question \textbf{P-RQ1}, the perceived speed of the capsule, the following values are aggregated: $M=1.93~m/s$, $SD = 0.998$, $Md = 1.95~m/s$, $Min = 0.87~m/s$, $Max = 3.25~m/s$. 
This represents an increase compared to the preset speed of $1.3~m/s$.
Regarding question \textbf{P-RQ2}, the preferred duration for the fade-to-black animation, the following results are gathered: $M=0.31~s$, $SD = 0.13~s$, $Md = 0.30~s$, $Min = 0.10~s$, $Max = 0.70~s$. 
In additional comments, three participants expressed a preference for a duration of $0~s$ if it was possible. 
One person finds a too-fast animation bothersome during frequent teleportation, while a long waiting time is also perceived as bothersome. 
Another person finds a too-long animation uncomfortable because it triggers blinking.
The participants who prefer a duration of $0~s$ also confirm an observation of impatience by Bozgeyikli et al \cite{Bozgeyikli2016PointAndTP}. 
As a result of this prestudy, we use a walking animation speed of $1.95~m/s$ and a fade-to-black animation time of $0.30~s$.

Parts of the prestudy were published as an extended abstract and presented as a poster at IEEE VR'24 \cite{woelwer2024posterFadeaway}.

\section{Study I: Distance Perception}\label{sec_study1}
In this first study, we want to evaluate the temporal influence of delayed teleportation methods on the perception of distances to answer research question RQ1. We therefore conduct an experiment employing a within-subject design, aimed at comparing distance perception with changing teleportation methods.

\subsection{Task}
The experiment's task design is based on a study by Keil et al. \cite{Keil}, which includes four key tasks: practicing locomotion, reaching and estimating a visible target, replicating a known distance, and training distance estimation.
All tasks occur on a straight path, providing clear movement direction and ensuring comparability under participants by requiring them to walk the same path.
As Peillard et al. \cite{Peillard2019ObjectsLookFarther} show, objects (e.g. target points) placed on the side of the user's field of view can be perceived to be farther away.
Therefore, having participants move the same straight path ensures comparability.
For our purposes, we exclude training in distance estimation, as it is not one of our goals.
Initially, participants are assigned a locomotion method and practice by reaching a nearby red target.
The primary task involves reaching a visible target between $90$ and $120$ meters from the start.
After reaching the target, participants estimate the distance and proceed to the next task, which involves replicating the just-traveled distance without a visible target.

We made adaptations with our integration of time proportional to distance in mind.
To address fatigue from covering long distances repeatedly, we reduced path lengths to $[30,70]~m$.
In the original study, longer distances were chosen to make the target initially invisible and thereby prevent purely visual estimates.
To keep this advantage, we visualize the target only when participants (or their avatars) are within the maximum teleportation range.
Additionally, intermediate points are irregularly spaced and visible only when participants are nearby, aiming to avoid counting strategies, i.e. participants trying to always cover equal steps of, e.g., $10~m$ in length.

The final task design includes a short practice phase, a distance estimation task, and a repetition task.
Each task begins with textual instructions presented through a sign in the virtual environment.
The practice phase familiarizes participants with the teleportation method by requiring them to reach an exemplary intermediate and final target point. 
In the main distance estimation task, participants progress through intermediate goals and finally the main target point.
Successfully reaching any target point is indicated by its removal and a short sound.
Possible placements of target points are presented in Table~\ref{tab_distPerc_pineconeDistances}.
Targets were chosen fixed per sequence of conditions and with overall similar distances covered per teleportation method amongst participants.
After reaching the target, participants estimate the distance by inputting their estimate in a virtual number field.
This concludes the task and immediately starts the next task at the user's current position.
This subsequent repetition task begins again with an information text. 
Participants must independently retrace the previous distance from the start to the final target point without intermediate goals or the target point itself visible.
They stop when they believe they have covered the distance and confirm their estimation.

\begin{table}
    \scriptsize
    \centering
    \caption{Possible options for the placement of intermediate goals (IG) and the final target point (Target) in meters from the starting point}
    \label{tab_distPerc_pineconeDistances}
    \begin{tabular}{@{}ccccc@{}}
    \toprule
    Option & IG 1   & IG 2   & IG 3   & Target   \\ \midrule
    1  & $15$ &       &       & $30$    \\
    2  & $13$ &       &       & $34$     \\
    3  & $13$ & $26$ &       & $39$     \\
    4  & $14$ & $28$ &       & $43$     \\ 
    5  & $15$ & $34$ &       & $47$ \\
    6  & $16$ & $29$ &       & $50$ \\
    7  & $13$ & $28$ &       & $50$  \\
    8  & $18$ & $34$ &       & $53$  \\
    9  & $14$ & $28$ & $42$ & $57$  \\
    10 & $15$ & $34$ & $47$ & $61$  \\
    11 & $16$ & $32$ & $50$ & $66$  \\
    12 & $18$ & $35$ & $49$ & $70$  \\
    \bottomrule
    \end{tabular}
\end{table}

\subsection{Virtual Environment}\label{sec_distPerc_virtEnv}
The virtual environment consists of a total path size of $650~m \times 5~m$. 
The ground is represented as a flat plane, with a $1~m$ wide cobblestone texture in the middle and grass textures on the left and right sides of the cobblestone (see Fig. \ref{img_distPerc_virtEnv_self}).
Objects are placed off the cobblestone path to improve reorientation after teleportation. 
The objects are chosen to resemble a forest walk theme.
The intermediate goals are visually matched to the theme. 
They are represented as oversized pine cones rotating on the vertical axis in the middle of the path. 
The final target point is represented by a red pillar, clearly distinct from the intermediate goals.
Proper object scaling is crucial for accurate distance perception. 
Therefore, the trees are set to a height of approximately $30~m$, which corresponds to the average tree height in Central Europe. 
Other objects, including branches on the ground, rocks, mushrooms, plants, and flying birds, are also scaled to average sizes, corresponding to the environment familiar to the locally recruited participants. 
To enhance immersion, forest sounds such as bird chirping are played.
Fig.~\ref{img_distPerc_virtEnv_self} depicts the starting point of the environment.

\subsection{Procedure}\label{sec_distPerc_procedure}
Participants are informed about the study and asked to sign a consent form.
Participants complete an initial demographic questionnaire before proceeding to the VR setup.
They are then instructed on how to wear the HMD and use the controller.
They initiate the distance perception experiment using one of the teleportation methods (experimental conditions) introduced in Section~\ref{sec_discPerc_tpm}. 
The order of conditions is balanced using Latin Square.
Each task is explained through integrated text, and in the initial repetitions, the experimenter provides additional verbal explanations to avoid misunderstandings.
For each teleportation method, there are three measurement repetitions.
After completing all tasks with one teleportation method, participants fill out a specific questionnaire related to that condition.
They then resume the VR setup for the next teleportation method, and the procedure is repeated for all teleportation methods.
Upon finishing the tasks with the fourth teleportation method, Study II (Section~\ref{sec_study2}) commences.
The total
average duration for both studies is $60$ minutes, with variations of $\pm15$ minutes depending on the participant.

\subsection{Measurements}\label{sec_distPerc_measurements}
To address the research questions, various measurements are taken during the experiment.
Central data of interest are the estimated distance $d_e$ and the distance to be estimated $d_t$.
For each participant, measurement repetition, task, and teleportation method we calculate a normalized error $e = \frac{d_e - d_t}{d_t}$.
The main measures for the analysis originate from formal descriptions of distance estimation errors that are described in the following.
To quantify participants' estimation performance, we use \textit{Bias}, \textit{Precision}, and \textit{Accuracy}.
\textbf{Bias} represents a systematic measurement error \cite{vocab}.
We quantify it using the mean normalized error (MPE \cite{Walther2005EstimationPerformance}) of the errors $e_1, e_2, e_3$ for the three measurement repetitions of a task a participant makes with a teleportation method.
\textbf{Precision}, which reflects the agreement between measurements obtained from repetitions of the same task \cite{vocab}, is measured using the Standard Deviation (SD) of the normalized errors \cite{Walther2005EstimationPerformance}.
\textbf{Accuracy}, describing the agreement between measured and true values \cite{vocab}, is quantified using the mean absolute normalized error (MAPE \cite{Walther2005EstimationPerformance}).

\subsection{Apparatus}\label{sec_distPerc_apparatus}
The study was implemented using the Unity game engine in version 2021.3.
Further, the \textit{XR Interaction Toolkit} package in version 2.3.1 was utilized including a basic version of instant teleportation.
The package was extended with the teleportation methods used in the experiments.
Hardware-wise, the study was conducted using the \textit{Meta Quest}.
Study participants were required to use one of the two corresponding controllers based on their dominant hand for interaction.

\subsection{Participants}\label{sec_distPerc_participants}
A total of $32$ participants ($19$ male, $13$ female) with ages ranging from $19$ to $64$ years ($M = 27.4$ years, $SD = 9.63$) were recruited.
Participants were reached through internal distribution channels at the University and the personal network of the experimenters.
Inclusion criteria for participation were normal or corrected vision and a minimum age of $18$ years.
Participants received compensation of $10$\texteuro.
Among the participants, $18$ reported prior experience with Virtual Reality, while $14$ reported having no prior experience.
Before the study, participants rated their ability to estimate physical distances on a seven-point scale from $1$ $\hat{=}$ ``Very poor'' to $7$ $\hat{=}$ ``Very good'', with an average of $M = 4.41$ ($SD = 1.24$). This data is for exploratory data analysis only and was not used for further analysis.

\subsection{Results}\label{sec_distPerc_results}

\begin{figure}
    \centering
    \includegraphics[width = 1\columnwidth]{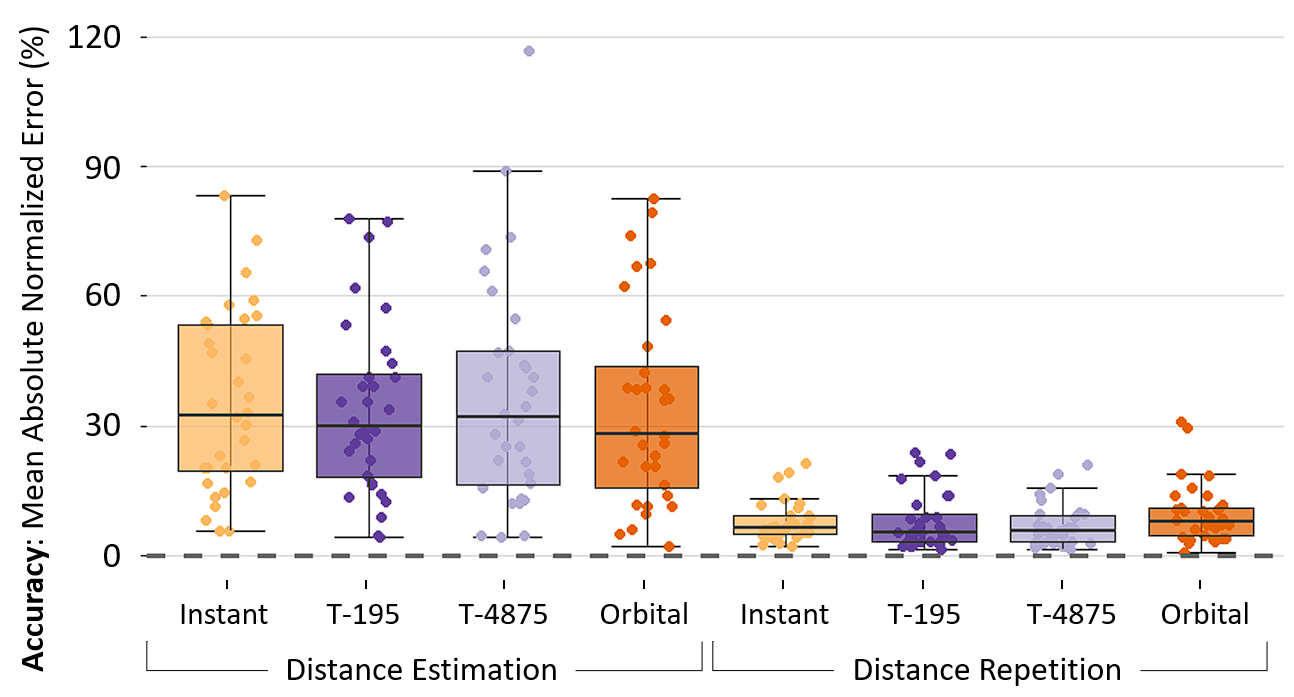}
    \vspace{-2.em}
    \caption{Mean Absolute Normalized Error (Accuracy) for all teleportation methods for the distance estimation and repetition task.\label{img_res_perf_MAE}}   
    \includegraphics[width = 1\columnwidth]{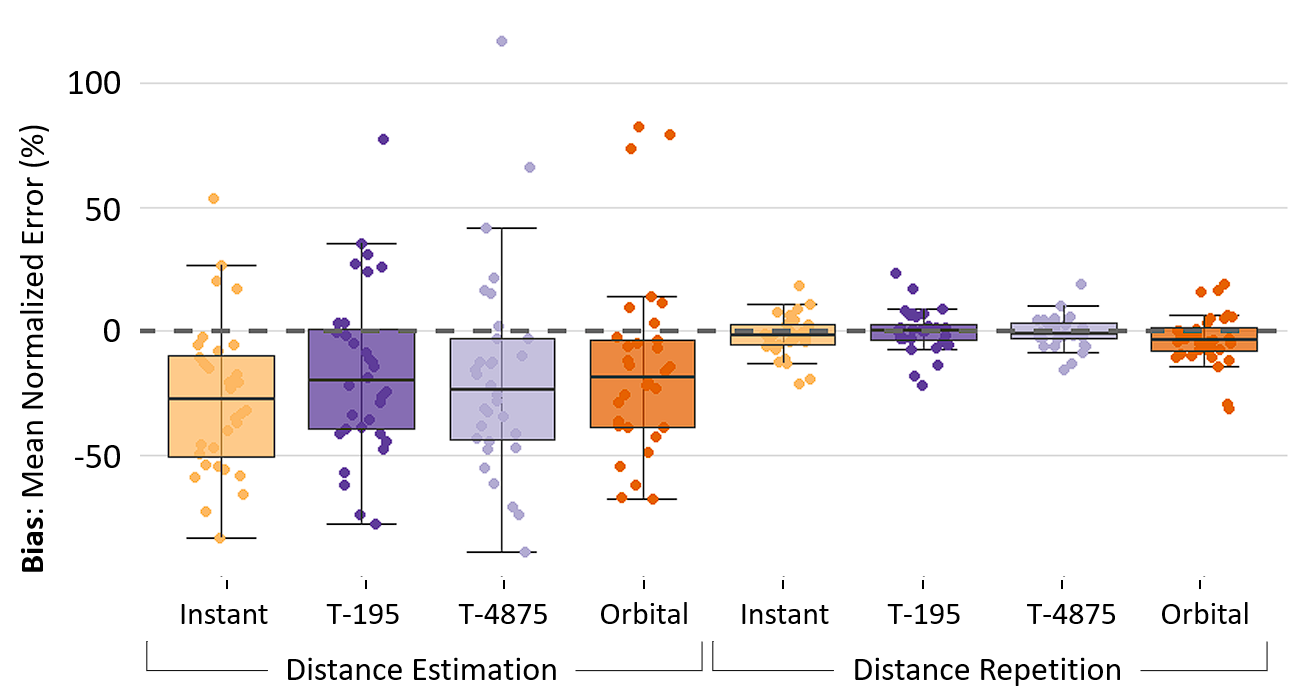}
    \vspace{-2.em}
    \caption{Mean Normalized Error (Bias) for all teleportation methods for the distance estimation and repetition task.\label{img_res_perf_ME}}  
    \includegraphics[width = 1\columnwidth]{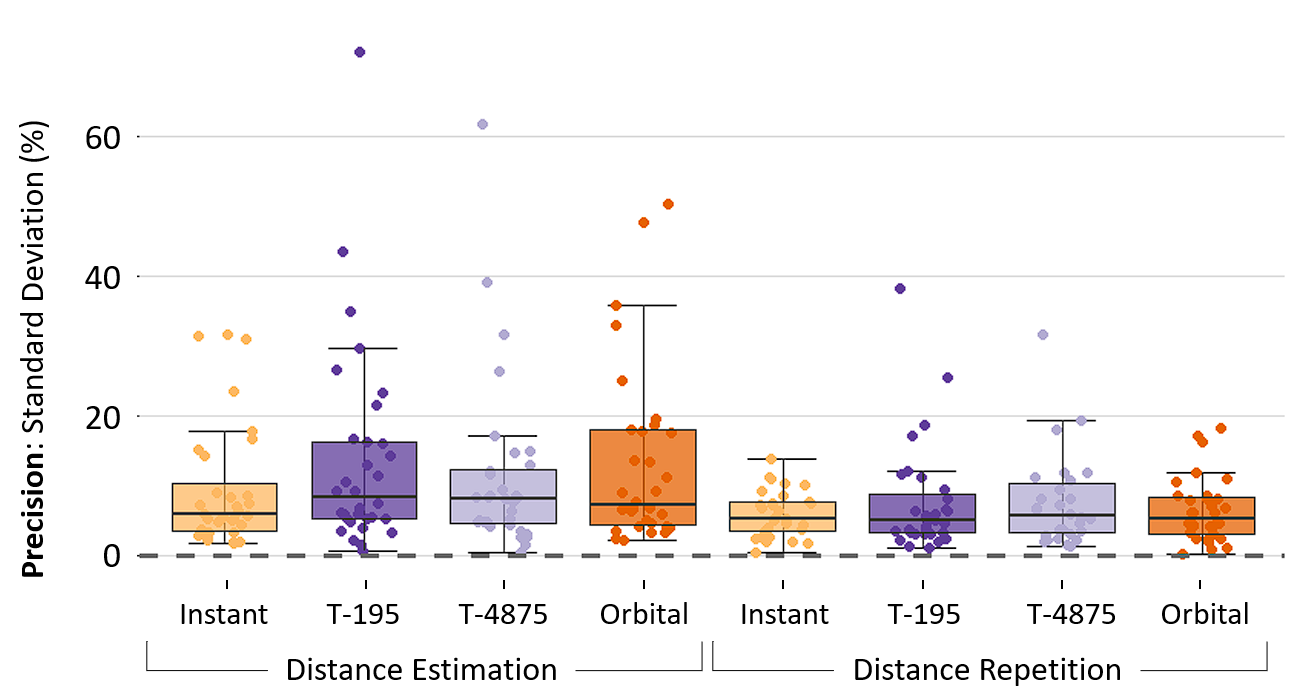}
    \vspace{-2.em}
    \caption{Standard deviation (Precision) of normalized errors for all teleportation methods for the distance estimation and repetition task.\label{img_res_perf_SD}}
\end{figure}

\input{tables/descriptives}
\input{tables/infTable}

The descriptive statistics of the distance estimation measures can be found in Table~\ref{tab_descr} and are visualized in Fig.~\ref{img_res_perf_MAE}-\ref{img_res_perf_SD}.
Section~\ref{sec_results_hypothesis} describes the hypotheses-based analysis of the collected data, while the following Section~\ref{sec_results_exploratory} describes an exploratory analysis.

\subsubsection{Hypotheses-based Analysis \label{sec_results_hypothesis}}
The inferential statistical analysis of the data is conducted using Paired-Sample t-tests.
In case the assumption of normality is violated or outliers are in the data, we use a non-parametric Wilcoxon signed rank test, instead. 
In case the data is also not symmetrical a Sign-Test is used. 
A significance level of $\alpha = .05$ is chosen for this analysis~\cite{fisher1956statistical}.
Effect sizes are assessed according to Cohen~\cite{cohen2013statistical}.
The complete inferential statistics are listed by hypotheses in Table~\ref{tab_inference}.

To test \textbf{H1}, which predicts a positive impact on distance estimation, when the teleportation is delayed, we test our three performance measures (Accuracy, Precision, and Bias) over the two experimental tasks: distance estimation and distance repetition.
To compensate for the test of multiple null hypotheses we use a Bonferroni correction and adjust the significance level to $\alpha = 0.0083$.
With the Bias significantly ($p = 0.008$) smaller in the delayed condition T1, we find a medium-sized correlation ($r=.43$) with fewer underestimations compared to the instantaneous method (IN).
We do not find significant effects in the other measures of Accuracy and Precision (see Table~\ref{tab_inference}, Hyp. 1), which we will further reflect on in the discussion.
Altogether we observe a significant improvement regarding Bias even after applying the Bonferroni correction without detecting adverse impacts on any other metric.
Thus, we argue that the delayed method contributes to an enhancement in distance estimation overall, validating H1.

\textbf{H2} predicts that a larger delay (T1 vs. T4) leads to systematically higher distance estimations.
We test the Bias over the two experimental tasks: distance estimation and distance repetition.
To compensate for the test of multiple null hypotheses we use a Bonferroni correction and adjust the significance level to $\alpha = 0.025$.
In both tasks, we do not find a difference (see Table~\ref{tab_inference}, Hyp. 2) and we thus cannot confirm H2.

\textbf{H3} predicts a difference in distance estimation performance when a delay happens in the target specification process with the user actively involved (OR) instead of the passive phase of pre-travel information (T1).
As in H1, we test our three performance measures.
To compensate for the test of multiple null hypotheses we use a Bonferroni correction and adjust the significance level to $\alpha = 0.0083$.
In both tasks, we do not find a difference (see Table~\ref{tab_inference}, Hyp. 3) and we thus cannot confirm H3.

\subsubsection{Exploratory Analysis \label{sec_results_exploratory}}
\input{tables/explAlternativeTable}
Based on our hypotheses, it made no sense to compare all the different methods with each other as they differ in multiple factors if they are not considered in pairs. 
However, such a comparison can be interesting depending on the situation and application. 
For this reason, we conclude an exploratory analysis of the collected data using a repeated measure ANOVA with the single factor, method (Instant, T-195, T-4875, Orbital).
We do assume the assumption of the normal sampling distribution to be met based on the central limit theorem given the sample size of N = 32 \cite{field2013discoveringSPSS}. 
In instances where extreme outliers are present, a non-parametric Friedman test is conducted instead. 
Sphericity is assessed using Mauchly's test. The Greenhouse–Geisser adjustment is applied to correct for any violations of sphericity when ($\epsilon > .75$).
Post-hoc tests are Bonferroni corrected.

The results are presented in Tabel~\ref{tab_exploratoryAnalysis}. 
A statistically significant difference is found for the performance measure of Bias ($F=(2.36, 73.2) = 3$, $p=.047$, $\eta^2 = .088$) in the distance estimation task. 
Other performance measures show no significant differences.

The pairwise post-hoc tests for Bias in the distance estimation task show no significant differences between teleportation methods: 
IN vs. T1: $M_{Diff} = -10.2$, 95\%-CI[$-22.3, 1.93$], $p=.15$, IN vs. T4: $M_{Diff} = -8.53$, 95\%-CI[$-23.4, 6.35$], $p=.70$, IN vs. OR: $M_{Diff} = -12.7$, 95\%-CI[$-28.3, 2.86$], $p=.17$, T1 vs. T4: $M_{Diff} = 1.67$, 95\%-CI[$-6.57, 9.91$], $p= 1.00$, T1 vs. OR: $M_{Diff} = -2.51$, 95\%-CI[$-13.6, 8.62$], $p=1.00$, T4 vs. OR: $M_{Diff} = -4.18$, 95\%-CI[$-16.9, 8.56$], $p=1.00$.

\section{Study II: Application}\label{sec_study2}
In the context of research question \textbf{RQ2}, a second study following a within-subject experimental design is conducted.
The aim is to investigate whether a potential effect of temporally delayed teleportation is transferred to the perception of space in virtual worlds.
This study is conducted in conjunction with and following Study I.
Therefore the apparatus (see Section \ref{sec_distPerc_apparatus}) and the participants (see Section \ref{sec_distPerc_participants}) are the same.
Regarding this experiment, users assessed their ability to estimate room sizes with an average of $M = 4.13$ ($SD = 1.39$) on a seven-point scale from $1$ $\hat{=}$ ``Very poor'' to $7$ $\hat{=}$ ``Very good''. Similarly to Study I this data is used for exploratory analysis only.

\subsection{Task}
The participants start in a room and have to reach a target point in an adjacent room.
They are prompted to freely explore both rooms.
Participants are encouraged by the experimenter to rely on intuitive estimation and reference objects provided, rather than counting the length of walls and calculating square meters.
Once participants feel they have obtained sufficient information, they must reach a target point to provide their estimation of the room combination size in square meters.
Upon entering the target point, participants are teleported to an estimation environment and prompted via a number field to provide their estimation in square meters.

\subsection{Virtual Environment}
\begin{figure}
    \centering
    \includegraphics[width=1.0\columnwidth]{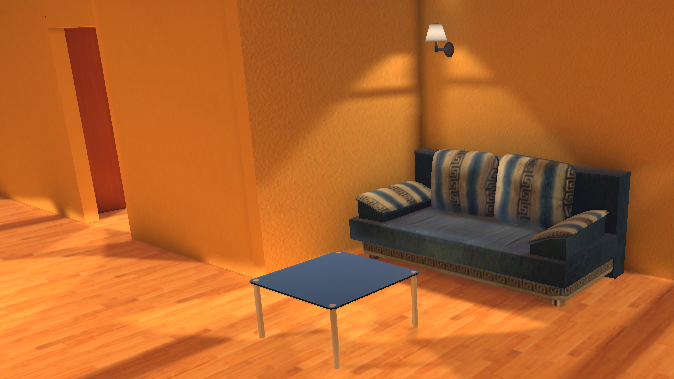}
    \caption{Virtual environment of the size estimation task. Exemplary view into room combination 1.}
    \label{img_application_virtEnv_room}
\end{figure}
\begin{figure}
    \centering
    \includegraphics[width = \columnwidth]{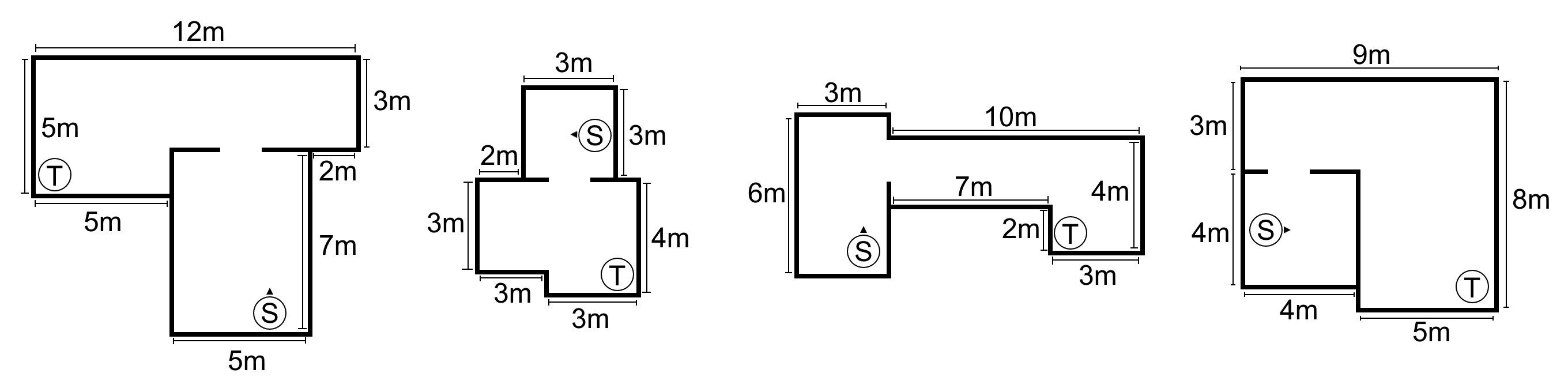}
    \caption{Room combinations used with sizes $81~m^2$(Room 1), $30~m^2$(Room 2), $44~m^2$(Room 3), $68~m^2$(Room 4) and their respective starting point (S) and target point (T). User rotation at the start of the task is indicated by an arrow on the starting point.}
    \label{img_application_virtEnv_floorPlans}
\end{figure}

The virtual environment consists of four differently sized independent pairs of rooms.
Each room combination consists of one rectangular-shaped and one L-shaped room.
One of the two rooms contains a starting point, while the second room contains a target point.
All rooms also contain furniture, including cabinets, shelves, beds, sofas, side tables, and lamps.
The walls are textured with wallpaper, while the floor is textured with laminate.
The furniture pieces and textures offer participants a reference size for estimating room dimensions and enhancing realism.
An overview of the floor plans of all rooms can be found in Fig.~\ref{img_application_virtEnv_floorPlans}.
The size of the room combinations is selected in the interval $[30,81]~m^2$, covering a range of various apartment sizes familiar to users.
An exemplary section of the interior of room combination 1 is shown in Fig.~\ref{img_application_virtEnv_room}.

\subsection{Procedure}\label{sec_application_procedure}
Participants use two of the familiar teleportation methods for this experiment, instant teleportation (Instant) and teleportation at walking speed (T-195) (see Section~\ref{sec_discPerc_tpm}).
The method order is balanced between participants. 
The estimation task is repeated twice for a teleportation method.
The sequence of rooms is balanced to provide different-sized combinations to users in different sequences of conditions.
Participants fill out a corresponding questionnaire after completing estimations with a teleportation method. 
They then perform the estimation task twice more using the second teleportation method and complete the questionnaire afterward.

\subsection{Results}
\begin{figure}
    \centering
    \includegraphics[width = \columnwidth]{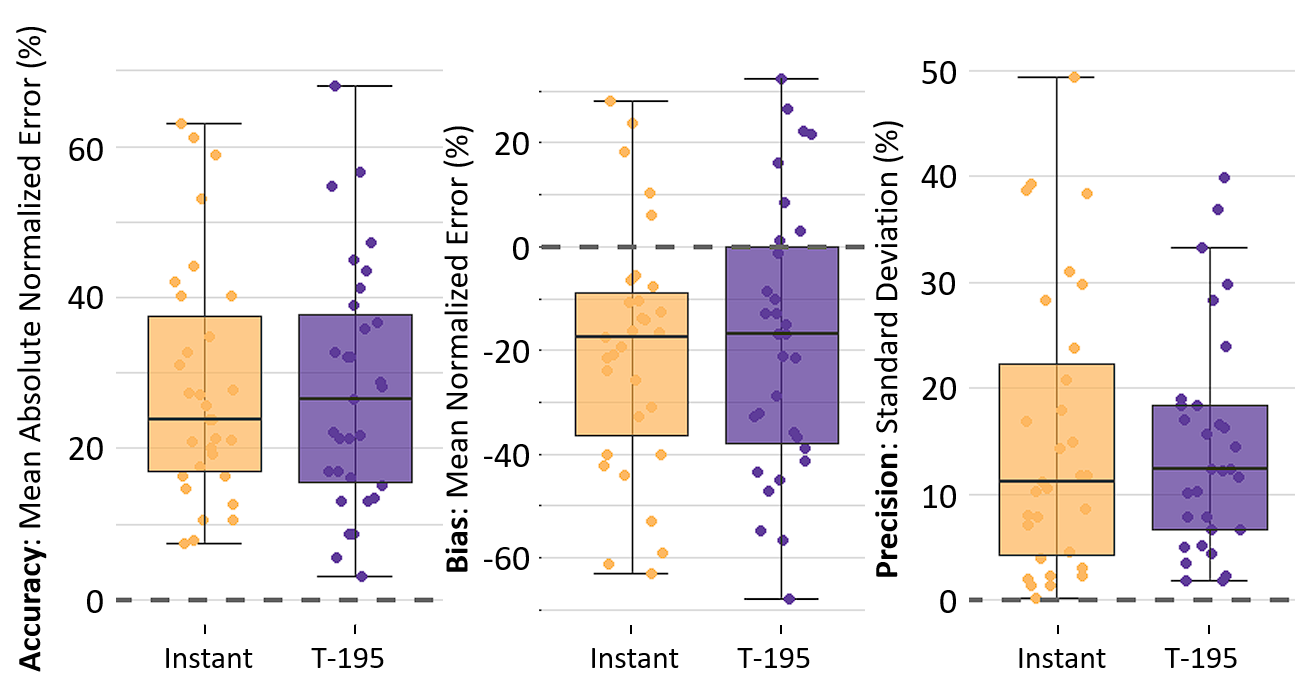}
    \vspace{-2.em}
    \caption{Performance Measures (Accuracy, Bias, and Precision) for both teleportation methods for the size estimation task.\label{img_res_perf_SE}}  
    \vspace{-1.em}
\end{figure}

The distance estimation measures from Study I (Accuracy, Precision, and Bias) are used analogously.
The descriptive statistics are visualized in Fig.~\ref{img_res_perf_SE}.
The inferential statistical analysis runs analogously to Study I.
The complete inferential statistics are listed by hypotheses in Table~\ref{tab_inference}.

To test \textbf{H4}, which predicts a positive impact on space estimation, when the teleportation is delayed, we test our three performance measures (Accuracy, Precision, and Bias) in the described room size estimation task.
To compensate for the test of multiple null hypotheses we use a Bonferroni correction and adjust the significance level to $\alpha = .017$.
Results show that for all three performance measures, users generally performed slightly better with the delayed teleportation method, manifesting in a smaller normalized absolute error (Accuracy), a smaller normalized underestimation (Bias), and more normalized precise estimations (Precision).
These improvements, however, turn out to be non-significant (see Table~\ref{tab_inference}, Hyp. 4) in any of the performance measures and we thus cannot confirm H4.

\section{Discussion}
In the following sections, various aspects of this work are discussed.
In Section~\ref{sec_disc_interpret_RQ1}, the collected results are interpreted with respect to research question RQ1, and in Section~\ref{sec_disc_interpret_RQ2}, the results regarding research question RQ2 are discussed.
Section~\ref{sec_disc_classification} places the results in the state of prior research.
Furthermore, in Section~\ref{sec_disc_reflection}, the methodology for evaluating the research question is reflected upon.
Finally, Section~\ref{sec_disc_limitations} discusses the limitations of the results of this work and considers their implications.

\subsection{Effects of delayed teleportation on distance perception}\label{sec_disc_interpret_RQ1}
\begin{table}
\scriptsize
\centering
\caption{Answers to the question ``Did you follow a specific strategy to solve the tasks? If yes, did the strategy differ depending on the teleportation method? If yes, how?'' during the final questionnaire split into categories \textit{visual}(V), \textit{temporal}(T), \textit{diverse}(D). Frequency of response in parentheses.}
\label{tab_strategies}
\begin{tabular}{cp{0.87\columnwidth}}
\toprule
 & Strategies                                                             \\ \midrule
\textbf{V} & Repetition of small, uniform distances ($21$)                               \\
 & Utilization of furniture as a reference size ($7$)                          \\
 & Orientation based on the size of environmental objects ($5$)                \\
 & Estimation of the length of the parabola ($3$)                              \\
 & Utilization of the target point indicator as a reference size ($2$)         \\
 & Estimation of the distance between the capsule and one's own position ($1$) \\
 & Utilization of the capsule as a reference size ($1$)                        \\
\textbf{T} & Counting the time of movement ($9$)                                         \\
 & Counting the meters during the capsule's movement ($2$)                     \\
 & Repetition of identical time intervals ($1$)                                \\
\textbf{D} & Unspecified differences between the teleportation methods ($3$)             \\
 & No difference between the teleportation methods ($2$)                       \\
 & Pure intuition ($1$)                                                        \\ \bottomrule
\end{tabular}
\vspace{-2em}
\end{table}

Regarding \textbf{H1}, a significantly improved performance in distance estimation can be observed through the performance measure of Bias.
This is a particularly important result as Bias is the only measure actively making a statement about the degree of underestimation.
Hence a significantly improved Bias is crucial to the further discussion of our research.
For the distance repetition task as well as Accuracy and Precision, no significant improvement can be observed. Performance stays relatively equal for these measures.
Based on these fairly similar performances but the significantly improved performance in Bias we argue, that in total an improved performance can be confirmed.
It is, however, evident, that improvement does not show in all the measures.
Our interpretation of these results is complex.
In general, there are two possible interpretations of the results we gather: (1) there is no difference between instantaneous and delayed teleportation regarding these measures or (2) the effect was covered by confounding factors.
While there not being an observable effect would be beneficial in arguing for instant teleportation as an easy and efficient way of travel, we believe that different strategies used to solve the tasks influence our results.
We had participants fill out questionnaires throughout the experiments, specifically after using each teleportation method and a concluding questionnaire, inquiring general feedback such as strategies used to solve the tasks or preferences for teleportation methods.
The data collected through these questionnaires may help in explaining.
Taking a look at the strategies listed in Table~\ref{tab_strategies} for solving the tasks shows that the most commonly used strategy is to repeat small, uniform distances.
One effect of this is that natural distance perception is bypassed, and the task is systematized.
This allows for more precise estimates.
Nearly two-thirds of participants report using this strategy.
The application of such a strategy was already suspected by Keil et al. \cite{Keil} but could not be confirmed through systematic questioning.
To better prevent such strategies, intermediate goals are placed in our experiment.
The effectiveness of this method therefore remains to be discussed.
Another reason for the lack of an effect is noted by Jones and Huang \cite{spacetime}.
In justifying Hypothesis H1, the Tau effect was mentioned.
Jones and Huang \cite{spacetime} point out that participants can consciously choose not to pay attention to the temporal context and can thus provide an unaffected estimate.
In the case of discrete target specification, this is especially conceivable since the destination point is chosen before any movement and thus any temporal influence occurs.
If participants estimate the distance already when selecting the destination point, they are not influenced by the time delay.

An effect regarding \textbf{H2} can not be demonstrated.
The assumption that a longer delay leads to a further estimation is reflected in both the distance estimation and replication task by a slightly higher mean estimation of distances with the walking speed method, but these differences turn out to be non-significant.
Similar to the justification for Hypothesis H1, counting strategies and avoidance of temporal influence through estimation before actual teleportation are conceivable here.
Another reason might be that the effect observed by Frenz and Lappe \cite{Frenz2005}, that a shorter duration of less than three seconds, as often occurs with the teleportation method at increased walking speed, only occurs in conjunction with an egocentrically perceived movement.
Since there is no movement of the users themselves during teleportation, but only of the preceding capsule, the effect may not be observable.
An exploratory comparison of the conditions T4 and IN reveals that the teleportation at increased walking speed does not significantly improve the Bias compared to instant teleportation, $z=1.44$, $p=.15$, $r=0.25$.

A confirming effect for \textbf{H3} is not observed.
A difference between the discrete and continuous target point selection methods only shows insignificant differences when considering Accuracy, Precision, and Bias.
Nevertheless, it is noticeable that the method with continuous target point selection (Orbital) tends to yield better results for the distance estimation task, while the method with discrete target point selection tends to yield better results for distance repetition.
The collected responses to the questionnaire in regards to the Orbital method show that even with the continuous target selection, users reported being able to reproduce short distances well and used slow movement to convert its duration into a metric value.
This suggests that counting strategies, as listed in Table~\ref{tab_strategies}, are also used here.
In addition to the most frequently reported strategy, repeating short distances, the answers suggest that the second most commonly chosen strategy, counting the duration of the movement, has an influence.
This also explains the almost identical values of both methods, as the capsules moved at the same speed and thus the duration of movement is the same. 

In conclusion, based on the significant reduction of underestimation with the delayed method as compared to the instant method, we argue that Hypothesis~H1 can be confirmed.
From this it concludes that there is an impact of delayed teleportation on perceived distances, answering \textbf{RQ1}.
By answering RQ1 we primarily show that time is a factor worth considering when a more verdical distance judgment is required.
This contributes a novel insight into the systematic investigation of underestimation in virtual reality and adds to the list provided by Creem-Regehr et al. \cite{CreemRegehr2022perceivingDist} where time is not yet considered.
These results, however, do not offer a general guideline on the usage of temporal delay for distance perception improvement.
Sub-questions \textbf{RQ1.1} and \textbf{RQ1.2}, cannot be answered conclusively.
While the performance in distance estimation significantly improved when using delayed teleportation, we do not see a difference among the delayed teleportation methods.
Generally, the delayed methods exhibit a lower tendency to underestimate than the immediate method, but show consistent absolute errors, indicating a propensity to overestimate instead.

\subsection{Effects of delayed teleportation on spatial perception}\label{sec_disc_interpret_RQ2}
The influence of temporally delayed teleportation methods on the perception of spatial dimensions is empirically evaluated in Study II (Section~\ref{sec_study2}).
The conducted experiment aims to address RQ2 -- asking for the impact on room size estimations.
Within the context of this research question, Hypothesis \textbf{H4} is formulated.
However, the results do not confirm the formulated hypothesis.

Hypothesis $4$ predicts an improvement in spatial size perception when employing a temporally delayed teleportation method.
The normalized absolute error values show a slightly improved estimation when using teleportation at walking speed, but these differences are found to be non-significant.
Similar characteristics are observable in the Bias and Precision results.
Participants using the delayed method exhibit a non-significant tendency to underestimate less and to be slightly more precise.
It can be argued that the lack of confirmation for hypothesis H4 is a continuation of the relatively equal performance measurements in hypothesis H1.
While we confirm H1, many of the recorded performance measures do not show significant improvement.
It is therefore feasible that an increasingly complicated task overshadows these improvements even more.

In addition, some qualitative results gathered through questionnaires indicate influencing factors, such as the parabolic ray from the controller to the target point, the target point indicator, and the size of the capsule.
Consequently, both methods possess aspects that participants used for orientation that are not solely attributable to the examined temporal aspect.
A central factor is also the participants' orientation to furniture as indicated in the strategy table.
While this orientation choice is intentional for external validity reasons, it possibly overshadows an effect of temporal modulation.
These findings also suggest that Research Question 2 cannot be answered conclusively and requires further investigations.

\subsection{Positioning of the results in the state of research}\label{sec_disc_classification}
A body of work has explored distance perception in VR, although fewer studies have focused on the impact of time.
Nevertheless, it is worthwhile to discuss the results in the context of existing research.

A commonality in our collected results across all teleportation methods is the tendency to underestimate distances.
This aligns with previous research findings.
Studies conducted by Waller and Richardson \cite{Waller2008CorrectingDistances} and Renner et al. \cite{Renner2013} revealed an average distance underestimation of approximately $29\%$ and $27\%$, respectively, when using HMDs.
In the case of instant teleportation, our results strongly align with prior research findings, showing an underestimation of approximately $27\%$. 
However, the results of the time-delayed teleportation methods represent an improvement in this tendency, with the T-195 method significantly reducing the underestimation to $16.9\%$ (and non-significant improvements to $14.3\%$ (Orbital) and $18.5\%$ (T-4875)).
This shows that while underestimation is not entirely compensated for, improvements of roughly $10\%$ are made.
To the best of our knowledge, this is the first time that such an improvement through the integration of time in teleportation interfaces has been shown.

The literature identifies different groups of influencing factors that affect distance perception and estimation \cite{Renner2013, CreemRegehr2022perceivingDist}.
Temporal influences and their effects can be classified into a group of human factors since the perception of time is a cognitive process.
Through the use of teleportation, two of the cues for accurate distance perception described by Creem-Regehr et al. \cite{CreemRegehr2022perceivingDist} are significantly limited. These are the factors of experiencing motion and the presence in the virtual world as cues for accurate distance perception.
Users using discrete teleportation methods feel less present in their environment, as observed by Bowman et al. \cite{Bowman1997TravelTechnique}.
Therefore, the fact that underestimations are still measured is not surprising and represents another manifestation of the studied misperceptions in virtual space.

Because the conducted experiment on distance perception was significantly inspired by that of Keil et al. \cite{Keil}, a comparison of the results is of particular interest.
However, it is important to note that the authors of the previous study used different target distance ranges.
Keil et al. report a mean absolute error (MAE) of $M=37.6~m$ $(SD = 26.8~m)$ for instant teleportation before the training phase and during the distance estimation task.
In our study, instant teleportation resulted in an MAE of $M=18.7~m$ $(SD=12.0~m)$.
The observed difference can be largely attributed to the distances being estimated.
The tendency to underestimate distances by around $27\%$ (among others, \cite{Renner2013}) leads to an expected MAE of $\frac{90+120}{2} \cdot 0.27 = 28.4~m$ for the target distance interval chosen by Keil et al. ($[90,120]~m$).
Based on the actual results, this corresponds to a normalized mean absolute error of $\frac{37.6}{28.4} = 1.324 = 32.4\%$.
This value closely matches our normalized absolute error value for the instant teleportation method of $35.3\%$.
It indicates that the use of instant teleportation produced nearly identical results in both experiments.
This similarity is surprising, especially considering the addition of objects in the virtual environment.
An environment-related context is often cited as a cue for more accurate distance estimation \cite{CreemRegehr2022perceivingDist, Renner2013}.
A comparison of the results of both experiments raises questions about this point.
One possible explanation is suggested by Murgia and Sharkey:
the authors find that underestimation of distances occurs similarly in both sparsely and densely equipped environments \cite{Murgia2009}.
For the distance replication task, Keil et al. report a MAE of $M=7.73~m$ $(SD=6.93~m)$ before training when using teleportation.
The experiment in our study yields an MAE of $M=8.20~m$ $(SD=4.29~m)$. 
It is notable that in both cases, distances are replicated much better than they were estimated in the distance estimation task.

\subsection{Reflection on the investigations carried out}\label{sec_disc_reflection}
The overall results show a high degree of consistency with research findings regarding instant teleportation.
The influence of a temporal component is only evident in a sub-scale of the performance measures used and there are complex reasons for this.
The absence of potential effects may also be attributed to the design of the study.
Therefore, a critical examination of the advantages and disadvantages of the experimental design is warranted.

Key aspects justifying a replication of the distance perception experiment by Keil et al. \cite{Keil} include the high reliability due to precise measurement of estimated and actual errors and the mandatory use of the locomotion method to solve the task.
A central drawback of the experiment, as suspected by the authors themselves, is the applicability of counting strategies.
When designing the experiment conducted in our work, special attention was therefore paid to limiting such tactics.
One solution to this is the irregular placement of intermediate goals.
Forcing participants to actively estimate differing distances gives us stronger results reflecting actual distance estimates.
The qualitatively evaluated results of the questionnaires suggest two conclusions in this regard: (1) the assumption made by Keil et al. \cite{Keil} regarding the application of counting tactics can be confirmed, and (2) the adjustments made in the experimental design could not entirely prevent their application.
Nevertheless, a reduction in their effectiveness can be assumed.
In the original experiment, participants used teleportation to estimate distances, fixing their controller's position to replicate a distance until they reached the target point. Only reaching the final target point required a different estimate.
Our intermediate goals require regular reorientation of the controller and a changed estimation.
Therefore, a new estimation of the distance is required frequently.
This represents an improvement in the experiment design.
It should be noted in every respect that the application of such strategies is difficult to avoid and is inherent to the nature of teleportation.
Just as steps can be counted when estimating distances in the real world, jumps can be counted in the virtual world when using teleportation.
To completely avoid this, there are two evaluation possibilities.
On the one hand, the characteristic target selection of teleportation could be changed (e.g. so that users cannot explicitly select a target point themselves but are rather teleported to a point selected by the system).
However, the practical relevance of such an investigation is questionable.
On the other hand, an experiment could be designed to measure distance perception without participants being aware of it.
This would also be much closer to the more application-relevant question of subconscious estimates.
How such an experiment would be designed remains open.

Potentially confounding to the results gathered could also be the environmental objects placed throughout the world.
As participants indicate in Table~\ref{tab_strategies} objects such as trees or branches were used during their execution of both studies to help them in estimating distances.
This was, however, intended as we see a need for the inclusion of such objects for multiple reasons.
One reason that applies to both the path and the room environment is the increase in external validity.
We want our results to be applicable in real-world applications rather than study environments and therefore decided on a more realistic environment.
Moreover, in the path environment, these objects allow for better comparability of the instant and the delayed methods.
While the delayed methods provide the user with an environmental distance cue, the virtual representative, the instant method is missing such a component.
Hence, in a sparsely equipped environment, the instant teleportation method would be disadvantaged and incentivize users to apply counting strategies since there are no other orientation cues.
Because of the intentional placement of the environmental objects, the frequency of people reporting to use them for distance estimation (Table~\ref{tab_strategies}) is in fact lower than what could be expected.
This, however, might come as a result of subconscious use or participants simply not reporting it.

\subsection{Limitations}\label{sec_disc_limitations}
Along with the results, we have already discussed a series of limitations, such as the issue of counting strategies. 
Furthermore, the visualization of temporal delay as an avatar-like representation through a moving capsule can have an impact on the experiment as well.
Creem-Regehr et al. \cite{CreemRegehr2022perceivingDist} identify experiencing an avatar as an influencing factor on the correct perception of distances.
The effect of using virtual representatives on the results, especially in the comparison of immediate and delayed teleportation, is not specifically examined in the conducted experiments.
Therefore, it cannot be ruled out that there are influences from this representation.
The objects placed in the environments behave similarly.
Renner et al. \cite{Renner2013} identify compositional factors, in addition to human factors, as influencing distance perception.
In the conducted experiments, the objects are intentionally placed with the aim of increasing external validity and comparability.
However, both investigations only aim to observe human perception.
The compositional factors arising from the objects are not considered.
Another limitation is the low number of repetitions per condition.
For a perceptual study, more trials would be beneficial.
Our first study repeats each task three times per teleportation method, while the second study is repeated twice per method.
Due to the number of teleportation methods and tasks, this was the maximum number of repetitions that could be fit within a 60-minute study.
However, a longer study could also induce fatigue in participants.
Lastly, our investigation into teleportation is highly motivated by the lack of cybersickness caused by it.
Comparing our delayed teleportation methods with steering techniques that implement a similar temporal component but allow for the perception of motion flow would be an interesting continuation of this work.
Other than the work by Keil et al. \cite{Keil} this would simulate an equal temporal component and therefore allow an investigation into other factors contributing to a differing distance estimation.

\section{Conclusion}
In summary, our research advances the understanding of distance misperception in VR, highlighting the influence of temporal factors.
Our findings demonstrate a reduction in underestimation of distances, from $27\%$ with instantaneous teleportation to $16.9\%$ with a time-delayed method, answering our primary research question of whether there is an influence of time on distance perception.
Since this temporal aspect remains largely unexplored in the literature, particularly in combination with teleportation, our studies contribute new insights into factors influencing distance underestimation.
Future research should therefore try to provide more universal guidelines on how to incorporate time to achieve a more verdical distance judgment.
Simultaneously, we are able to corroborate previous research findings regarding the underestimation observed with instantaneous teleportation specifically using modern hardware.
Regarding our second research question on the impact of time-delayed methods on room size estimations, the results remain inconclusive.
Despite improving upon previous task designs, our study cannot entirely eliminate counting strategies, simplifying distance estimation with teleportation.
However, systematic questioning confirms suspicions raised in prior research regarding their usage.
Lastly, our pre-study investigates the preferred duration for fade-to-black transitions, settling at $0.3~s$, and adds understanding of exocentrically perceived walking speed in virtual environments which participants judge to be $1.95~m/s$.

\bibliographystyle{IEEEtran}
\bibliography{bib_23-06-05}

%\begin{IEEEbiography}[{\includegraphics[width=1in,height=1.25in,clip,keepaspectratio]{figures/Ashu.jpg}}]{Ashu Adhikari}
% has a Master's degree from the School of Interactive Arts and Technology (SIAT), Simon Fraser University (SFU). As an MSc student he joined the iSPACE team and has collaborated ever since. He pursues his interest in embodied locomotion interfaces in virtual reality.
%\end{IEEEbiography}
%\vskip -2\baselineskip plus -1fil
%\begin{IEEEbiography}[{\includegraphics[width=1in,height=1.25in,clip,keepaspectratio]{figures/daniel.png}}]{Daniel Zielasko}
%is currently a postdoctoral researcher and lecturer at the University of Trier’s HCI group. He received his doctoral degree in 2020 at the Virtual Reality and Immersive Visualization group at RWTH Aachen University for studying desk-centered virtual reality. He has worked together with neuroscientists, psychologists, medical technicians, and archaeologists on projects such as the EU flagship project HBP (Human Brain Project) and the SMHB (Supercomputing and Modeling for the Human Brain). Today, he is working on the integration of VR technologies and methods into everyday life, such as existing professional workflows and entertainment.
%\end{IEEEbiography}
%\vskip -2\baselineskip plus -1fil

\end{document}

%% file: tables/descriptives.tex
\begin{table*}
\caption{Descriptive statistics (M = mean, SD = standard deviation) for each measure and task. \label{tab_descr}}
\centering
\scriptsize
\begin{tabular}{llllllllllllllllll}
\hline
          && \multicolumn{8}{c}{Distance Estimation (DT)}                                                                            & \multicolumn{8}{c}{Distance Repetition (RT)}                                                                            \\
\hline
          && \multicolumn{2}{c}{Instant} & \multicolumn{2}{c}{T-195} & \multicolumn{2}{c}{T-4875} & \multicolumn{2}{c}{Orbital} & \multicolumn{2}{c}{Instant} & \multicolumn{2}{c}{T-195} & \multicolumn{2}{c}{T-4875} & \multicolumn{2}{c}{Orbital} \\
          && M            & SD           & M           & SD          & M            & SD          & M             & SD          & M             & SD          & M           & SD          & M            & SD          & M            & SD           \\
\hline
\rowcolor[HTML]{EFEFEF}Accuracy  && 35.3        & 20.8        & 33.9       & 20.0       & 36.5        & 26.1       & 33.9         & 22.8       & 7.91          & 4.88        & 7.89        & 6.70        & 6.98         & 5.05        & 8.96          & 6.17         \\
Bias      && -27.0        & 30.5        & -16.8       & 34.2       & -18.5       & 41.1       & -14.3        & 37.5       & -1.96         & 8.41        & 0.39         & 8.68        & -0.19         & 6.57        & -2.21         & 9.75         \\
\rowcolor[HTML]{EFEFEF} Precision && 9.40          & 8.82         & 13.8        & 14.8       & 11.4        & 12.6       & 13.2         & 12.7       & 6.00             & 3.35        & 7.69        & 8.20         & 7.59         & 6.61        & 6.62         & 4.72        \\
\hline
\end{tabular}
\end{table*}

%% file: tables/infTable.tex
\begin{table*}
\caption{Pairwise (Condition A vs. B) inferential statistical analysis of the hypotheses, where Hyp. specifies the hypothesis number, Task (DT = distance estimation, RT = repetition task, SE = space estimation),  DV the dependent variable, Dir. denotes the direction of the hypothesis, followed by the t or Z statistics. In the latter case, i.e., when a non-parametric test was performed, SDs and df are not specified.  S.-Wilk reports the p-value for the Shapiro-Wilk test regarding normally distributed residuals.\label{tab_inference}}
\centering
\scriptsize
\begin{tabular}{ccclcrcccrcrrcrr}
\textbf{Pair} &
  \textbf{Hyp.} &
  \textbf{Task} &
  \textbf{DV} &
  \textbf{Cond. A} &
  \multicolumn{1}{l}{\textbf{M / Mdn}} &
  \multicolumn{1}{c}{\textbf{SD}} &
  \multicolumn{1}{l}{\textbf{Dir.}} &
  \textbf{Cond. B} &
  \multicolumn{1}{l}{\textbf{M / Mdn}} &
  \multicolumn{1}{c}{\textbf{SD}} &
  \multicolumn{1}{l}{\textbf{S.-Wilk}} &
  \multicolumn{1}{l}{\textbf{t / Z}} &
  \multicolumn{1}{l}{\textbf{df}} &
  \multicolumn{1}{c}{\textbf{p}} &
  \multicolumn{1}{l}{\textbf{d / r}} \\
\rowcolor[HTML]{EFEFEF} 
1 &
  \cellcolor[HTML]{EFEFEF} &
  DT &
  Accuracy &
  \cellcolor[HTML]{EFEFEF} &
  35.3 &
  20.8 &
  \textgreater{} &
  \cellcolor[HTML]{EFEFEF} &
  33.9 &
  20.0 &
  .12 &
  0.39 &
  31 &
  .35 &
  0.069 \\
\rowcolor[HTML]{EFEFEF} 
2 &
  \cellcolor[HTML]{EFEFEF} &
  \cellcolor[HTML]{EFEFEF}DT &
  $\mid$Bias$\mid$ &
  \cellcolor[HTML]{EFEFEF} &
  $\mid$-27.5$\mid$ &
  - &
  \textgreater{} &
  \cellcolor[HTML]{EFEFEF} &
  $\mid$-20.1$\mid$ &
  - &
  .071 &
  2.41 &
  - &
  \textbf{.008} &
  0.43 \\
\rowcolor[HTML]{EFEFEF} 
3 &
  \cellcolor[HTML]{EFEFEF} &
  \cellcolor[HTML]{EFEFEF}DT &
  Precision &
  \cellcolor[HTML]{EFEFEF} &
  5.83 &
  - &
  \textgreater{} &
  \cellcolor[HTML]{EFEFEF} &
  8.35 &
  - &
  \textless~.001 &
  -1.81 &
  - &
  .97 &
  0.32 \\
\rowcolor[HTML]{EFEFEF} 
4 &
  \cellcolor[HTML]{EFEFEF} &
  RT &
  Accuracy &
  \cellcolor[HTML]{EFEFEF} &
  6.51 &
  - &
  \textgreater{} &
  \cellcolor[HTML]{EFEFEF} &
  5.26 &
  - &
  .063 &
  0.22 &
  - &
  .42 &
  0.039 \\
\rowcolor[HTML]{EFEFEF} 
5 &
  \cellcolor[HTML]{EFEFEF} &
  RT &
  $\mid$Bias$\mid$ &
  \cellcolor[HTML]{EFEFEF} &
  $\mid$-1.96$\mid$ &
  8.41 &
  \textgreater{} &
  \cellcolor[HTML]{EFEFEF} &
  $\mid$0.39$\mid$ &
  8.68 &
  .20 &
  1.40 &
  29 &
  .090 &
  0.26 \\
\rowcolor[HTML]{EFEFEF} 
6 &
  \multirow{-6}{*}{\cellcolor[HTML]{EFEFEF}1} &
  RT &
  Precision &
  \multirow{-6}{*}{\cellcolor[HTML]{EFEFEF}IN} &
  5.21 &
  - &
  \textgreater{} &
  \multirow{-6}{*}{\cellcolor[HTML]{EFEFEF}T1} &
  5.11 &
  - &
  \textless~ .001 &
  0 &
  - &
  .50 &
  0 \\
7 &
   &
  DT &
  Bias &
   &
  -16.8 &
  34.2 &
  \textgreater{} &
   &
  -18.5 &
  41.1 &
  .080 &
  0.57 &
  31 &
  .29 &
  0.10 \\
8 &
  \multirow{-2}{*}{2} &
  RT &
  Bias &
  \multirow{-2}{*}{T1} &
  -0.71 &
  8.91 &
  \textgreater{} &
  \multirow{-2}{*}{T4} &
  -0.38 &
  6.54 &
  .42 &
  -0.18 &
  30 &
  .43 &
  0.032 \\
\rowcolor[HTML]{EFEFEF} 
9 &
  \cellcolor[HTML]{EFEFEF} &
  DT &
  Accuracy &
  \cellcolor[HTML]{EFEFEF} &
  30.0 &
  - &
   $\neq$ &
  \cellcolor[HTML]{EFEFEF} &
  28.2 &
  - &
  .003 &
  0.62 &
  - &
  .54 &
  0.11 \\
\rowcolor[HTML]{EFEFEF} 
10 &
  \cellcolor[HTML]{EFEFEF} &
  DT &
  $\mid$Bias$\mid$ &
  \cellcolor[HTML]{EFEFEF} &
  $\mid$-20.2$\mid$ &
  - &
   $\neq$ &
  \cellcolor[HTML]{EFEFEF} &
  $\mid$-18.5$\mid$ &
  - &
  .010 &
  0.43 &
  - &
  .67 &
  0.076 \\
\rowcolor[HTML]{EFEFEF} 
11 &
  \cellcolor[HTML]{EFEFEF} &
  DT &
  Precision &
  \cellcolor[HTML]{EFEFEF} &
  8.35 &
  - &
   $\neq$ &
  \cellcolor[HTML]{EFEFEF} &
  7.26 &
  - &
  .039 &
  0.17 &
  - &
  .87 &
  0.029 \\
\rowcolor[HTML]{EFEFEF} 
12 &
  \cellcolor[HTML]{EFEFEF} &
  RT &
  Accuracy &
  \cellcolor[HTML]{EFEFEF} &
  7.96 &
  6.60 &
   $\neq$ &
  \cellcolor[HTML]{EFEFEF} &
  9.65 &
  7.00 &
  .20 &
  -1.16 &
  31 &
  .26 &
  0.21 \\
\rowcolor[HTML]{EFEFEF} 
13 &
  \cellcolor[HTML]{EFEFEF} &
  RT &
  $\mid$Bias$\mid$ &
  \cellcolor[HTML]{EFEFEF} &
  $\mid$-0.20$\mid$ &
  8.79 &
   $\neq$ &
  \cellcolor[HTML]{EFEFEF} &
  $\mid$-3.32$\mid$ &
  10.7 &
  .23 &
  -1.58 &
  31 &
  .13 &
  0.28 \\
\rowcolor[HTML]{EFEFEF} 
14 &
  \multirow{-6}{*}{\cellcolor[HTML]{EFEFEF}3} &
  RT &
  Precision &
  \multirow{-6}{*}{\cellcolor[HTML]{EFEFEF}T1} &
  5.11 &
  - &
   $\neq$ &
  \multirow{-6}{*}{\cellcolor[HTML]{EFEFEF}OR} &
  5.34 &
  - &
  .001 &
  -0.94 &
  - &
  .35 &
  0.17 \\
15 &
   &
  SE &
  Accuracy &
   &
  28.2 &
  15.6 &
  \textgreater{} &
   &
  27.9 &
  16.1 &
  .051 &
  0.14 &
  30 &
  .45 &
  0.025 \\
16 &
   &
  SE &
  $\mid$Bias$\mid$ &
   &
  $\mid$-17.5$\mid$ &
  - &
  \textgreater{} &
   &
  $\mid$-17.0$\mid$ &
  - &
  .28 &
  0.73 &
  - &
  .23 &
  0.13 \\
17 &
  \multirow{-3}{*}{4} &
  SE &
  Precision &
  \multirow{-3}{*}{IN} &
  15.2 &
  13.4 &
  \textgreater{} &
  \multirow{-3}{*}{T1} &
  14.5 &
  10.4 &
  .073 &
  0.24 &
  30 &
  .41 &
  0.042
\end{tabular}
\end{table*}

%% file: tables/explAlternativeTable.tex
\begin{table}
\caption{Summary of Exploratory Data Analysis: DV denotes the dependent variable, Task (DT = distance estimation, RT = repetition task), $df_b$ the between-groups degrees of freedom. In cases where a Repeated Measures ANOVA was performed, $df_w$ denotes the within-groups degrees of freedom. The value - in this column indicates that a Friedman Test was performed instead. F / $\chi^2$ denotes the F-statistic when an ANOVA was performed or the Chi-squared statistic when a Friedman Test was performed. The p-value is given for the respective test. Effect sizes $\eta_p^2$ are given when an ANOVA was performed and Kendall's W respectively when a Friedman test was performed.\label{tab_exploratoryAnalysis}}
\centering
\scriptsize
\begin{tabular}{@{}cccccccc@{}}
\toprule
Pair                        & DV            & Task & $df_b$     & $df_w$      & $F$ / $\chi^2$ & $p$         & $\eta_p^2$ / $W$ \\ \midrule
\rowcolor[HTML]{EFEFEF} 1   & Accuracy      & DT   & 3.00       & 93.0      & 0.24           & .89           & 0.008\\
                        2   & Bias          & DT   & 2.36       & 73.2      & 3.00           & \textbf{.047} & 0.088\\
\rowcolor[HTML]{EFEFEF} 3   & Precision     & DT   & 3.00       & -         & 4.39           & .22           & 0.046\\
                        4   & Accuracy      & RT   & 3.00       & 87.0      & 0.84           & .47           & 0.028\\
\rowcolor[HTML]{EFEFEF} 5   & Bias          & RT   & 3.00       & 87.0      & 1.05           & .37           & 0.035\\
                        6   & Precision     & RT   & 3.00       & -         & 1.10           & .78           & 0.013\\ \bottomrule
\end{tabular}
\end{table}